\documentclass[preprint,showpacs,preprintnumbers,amsmath,amssymb,superscriptaddress]{revtex4}

\include{ams}
\usepackage{amsmath,amssymb,amsthm}
\usepackage{dcolumn}
\usepackage{bm}
\usepackage{graphicx}

\begin{document}

\title{Edge states of graphene bilayer strip }

\author{M. Pudlak}\email{pudlak@saske.sk}
\author{R. Pincak}\email{pincak@saske.sk}
\affiliation{Institute of Experimental Physics, Slovak Academy of
Sciences, Watsonova 47,043 53 Kosice, Slovak Republic}

\date{\today}


\begin{abstract}
The electronic structure of the zig-zag bilayer strip is analyzed. The
electronic spectra of the bilayer strip is computed. The
dependence of the edge state band flatness on the bilayer width is
found. The density of states at the Fermi level is
analytically computed. It is shown that it has the singularity
which depends on the width of the bilayer strip. There is also
asymmetry in the density of states below and above the Fermi energy.

\end{abstract}

\maketitle

\section{Introduction}

Carbon atoms can create a variety of forms such a graphite,
diamond, carbon fibers, fullerenes and carbon nanotubes. A carbon
nanotube can be described as a graphene sheet rolled into a
cylindrical shape so that the structure is one-dimensional with
axial symmetry and in general exhibiting a spiral conformation
called chirality. They are interesting  because of their unique
mechanical and electronic properties~\cite{Saito}. From the
pioneering works ~\cite{Wal,Carter,Slon}, the electronic properties
of graphite have attracted interest because of unconventional
physical properties of a graphite layer. The development in the
fabrication of the single layers of graphite
(graphene)~\cite{Novoselov} caused a striking level of interest
in the investigation of the carbon compositions. In addition to the
closed carbon molecules ~\cite{pudlak}, systems with boundaries also
show interesting features. The nanographite zig-zag ribbon possesses
localized edge states near the Fermi level. States like that are
absent for ribbons with armchair edges~\cite{Fujita}. The graphite sheet
is a zero-gap semiconductor with the density of states(DOS)
vanishing at the Fermi level, the edge states of the zig-zag ribbons
produce a peak in the DOS at the Fermi level. Both the carbon
nanotubes and graphite layers have the
edge states because of their boundary. The presence of the edge state results in the
relatively important contribution to the density of states (DOS)
near the Fermi energy. It was found ~\cite{Hod,NOVA} that the
HOMO-LUMO (highest occupied molecular orbital and lowest unoccupied
molecular orbital,respectively) gap is inversely proportional to the
length of the zig-zag carbon nanotube segment. The zig-zag ribbons
have partly flat bands at the Fermi level~\cite{Fujita,Li}. In the
presented paper, we focus on the computation of the electronic
spectra of the zig-zag bilayer strip and also on the computation of
the DOS of the edge states near the Fermi level.

\section{Theory}

\begin{figure}[htb]
\centering
\includegraphics[width=0.8\textwidth]{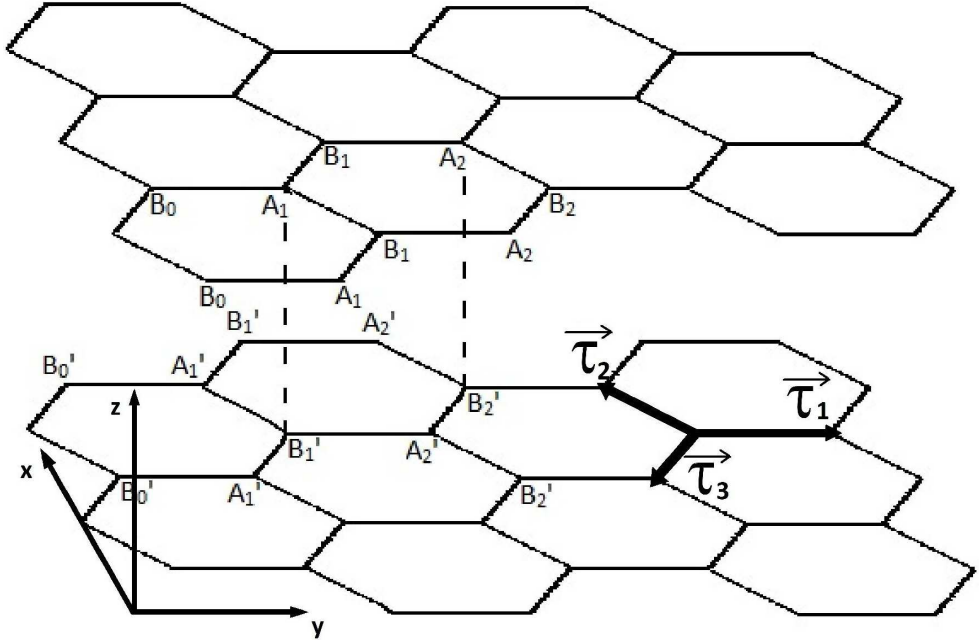}
\caption{Structure of finite-width graphene bilayer. }\label{fig14}
\end{figure}

Firstly, we describe the model for the zig-zag bilayer strip. We will
study the edge and size effects using the tight-binding model for
this strip shown in Fig.~\ref{fig14}.

The $\pi $ electronic structures are calculated from the
tight-binding Hamiltonian
$$ H=\sum_{i}\epsilon_{i} |\varphi
^{u}_{i}\rangle \langle \varphi ^{u}_{i}| +\sum _{i,j}\gamma_{ij}
\left(|\varphi ^{u}_{i}\rangle \langle \varphi ^{u}_{j}|+h.c\right)+
\sum_{i} \widetilde{\epsilon_{i}}|\varphi ^{d}_{i}\rangle \langle
\varphi ^{d}_{i}|+\sum _{i,j}\widetilde{\gamma}_{ij} \left(|\varphi
^{d}_{i}\rangle \langle \varphi ^{d}_{j}|+h.c\right) $$
\begin{equation}
+\sum_{l,n} W_{ln} \left(|\varphi ^{d}_{l}\rangle \langle \varphi
^{u}_{n}|+h.c\right),
\end{equation}
$\epsilon_{i}$ and $\widetilde{\epsilon_{i}}$ are the site energies
of the upside and down layer; $|\varphi ^{u}_{i}\rangle$, $|\varphi
^{d}_{i}\rangle$ are $\pi $ the orbitals on site $i$ at the upside and
down layer; $\gamma_{ij}$, $\widetilde{\gamma}_{ij}$ are the
intralayer hopping integrals; $W_{ij}$ are the interlayer hoping
integrals which depend on the distance $d_{ij}$ and angle
$\theta_{ij}$ between the $\pi_{i}$ and $\pi_{j}$ orbitals.

To describe the parameter which characterizes the zig-zag bilayer
strip, we start from the graphene layer where we can define the
vectors connecting the nearest neighbor carbon atoms in the form:
\begin{equation}
\overrightarrow{\tau_{1}}=a(0;\frac{1}{\sqrt{3}}),\nonumber
\end{equation}
\begin{equation}
\overrightarrow{\tau_{2}}=a(\frac{1}{2};-\frac{1}{2\sqrt{3}}),\nonumber
\end{equation}
\begin{equation}
\overrightarrow{\tau_{3}}=a(-\frac{1}{2};-\frac{1}{2\sqrt{3}}).
\end{equation}

The distance between atoms in the unit cell is
$d=|\overrightarrow{\tau_{i}}|=\frac{a}{\sqrt{3}}$. We want to find
a solution to the double-layer graphene strip in the form:
\begin{equation}
\psi(\overrightarrow{r})=\psi^{u}(\overrightarrow{r})+\psi^
{d}(\overrightarrow{r})
\end{equation}
where
\begin{eqnarray}
\psi^{u}(\overrightarrow{r})=\sum_{i=0}^{M+1}\left(C_{A_{i}}\psi_{A_{i}}
+C_{B_{i}}\psi_{B_{i}}\right),
\end{eqnarray}
and
\begin{equation}
\psi^{d}(\overrightarrow{r})=\sum_{i=0}^{M+1}\left(C_{A_{i}^{`}}\psi_{A_{i}^{`}}+C_{B_{i}^{`}}\psi_{B_{i}^{`}}\right).
\end{equation}
We want to find a solution to the above equation in the form of the
Bloch function
\begin{equation}
\psi_{\alpha}(\overrightarrow{k},\overrightarrow{r})=
\frac{1}{\sqrt{N}}\sum_{n}e^{i\overrightarrow{k}.\overrightarrow{r_{n}}
} |\varphi(\overrightarrow{r}-\overrightarrow{r}_{n})\rangle,
\end{equation}
where $\alpha$ denotes $A$ or $B$ atoms. Here
$\overrightarrow{k}=(k_{x},0)$, $\overrightarrow{r_{n}}$ is the
position of a unit cell and $N$ is the number of a unit cell;
$|\varphi(\vec{r})\rangle$ is a $\pi$ orbital. We denote
\begin{equation}
\epsilon_{i}=\widetilde{\epsilon}_{i}=\Delta=
\langle\varphi^{out}(r-A_{i})|H|\varphi^{out}(r-A_{i})\rangle=
\langle\varphi^{out}(r-B_{i}^{'})|H|\varphi^{out}(r-B_{i}^{'})\rangle,
\end{equation}
\begin{equation}
\epsilon_{i}=\widetilde{\epsilon}_{i}=-\Delta=
\langle\varphi^{out}(r-B_{i})|H|\varphi^{out}(r-B_{i})\rangle=
\langle\varphi^{out}(r-A_{i}^{'})|H|\varphi^{out}(r-A_{i}^{'})\rangle.
\end{equation}
Now we define the intratube hopping integrals as
$\gamma_{ij}=\tilde{\gamma}_{ij}=\gamma_{0}$. We take into account
only the interaction between nearest-neighbors also in the case of
interlayer interaction and denote
\begin{equation}
\langle\varphi(r-A_{i})|H|\varphi(r-B_{i}^{'})\rangle=\gamma_{1}.
\end{equation}
In confining the structure along the width, the edge states are
induced by terminating the width dimension with zig-zag shaped
edges. The presence of edges in the bilayer strip changes the
dimensionality of the system from a two-dimensional to a one-dimensional
system. The electronic spectrum of the zig-zag bilayer strip can be
described by the following system of equations:
\begin{equation}
\label{4.1}
(E-\Delta)C_{A_{m}}=-\gamma_{0}C_{B_{m-1}}-g_{k}C_{B_{m}}-\gamma_{1}C_{B'_{m}},
\end{equation}
\begin{equation}
\label{4.2}
(E+\Delta)C_{B_{m}}=-\gamma_{0}C_{A_{m+1}}-g_{k}C_{A_{m}},
\end{equation}
\begin{equation}
\label{4.3}
(E+\Delta)C_{A'_{m}}=-\gamma_{0}C_{B'_{m-1}}-g_{k}C_{B'_{m}},
\end{equation}
\begin{equation}
\label{4.4}
(E-\Delta)C_{B'_{m}}=-\gamma_{0}C_{A'_{m+1}}-g_{k}C_{A'_{m}}-\gamma_{1}C_{A_{m}},
\end{equation}
where
\begin{equation}
g_{k}=2 \gamma_{0} \cos(k_{x}a/2).
\end{equation}
Here $m=1,.....,M$, are site indices, where $M$ describes the width
of the graphene bilayer.

We assume that the $A_{0}$ and $B_{M+1}$ sites are missing. So we
have the boundary condition
$C_{A_{0}}=C_{B_{M+1}}=C_{A'_{0}}=C_{B'_{M+1}}=0$. The solution is
assumed to be~\cite{Wak}
\begin{equation}
C_{A_{m}}=Ae^{ipm}+Be^{-ipm},
\end{equation}
\begin{equation}
C_{B_{m}}=Ce^{ipm}+De^{-ipm}.
\end{equation}
Here $A,B,C$ and $D$ are the coefficients which have to be determined,
and $p$ is the transverse wave number. From the boundary condition
we have
\begin{equation}
C_{A_{0}}=A+B=0,
\end{equation}
\begin{equation}
C_{B_{M+1}}=Ce^{ip(M+1)}+De^{-ip(M+1)}=0.
\end{equation}
And so
\begin{equation}
\label{4.11} C_{A_{m}}=A(e^{ipm}-e^{-ipm}),
\end{equation}
\begin{equation}
\label{4.12} C_{B_{m}}=C(e^{ipm}-z^{2}e^{-ipm}),
\end{equation}
where $z=e^{ip(M+1)}$. And similarly,

\begin{equation}
\label{4.11a} C_{A'_{m}}=A'(e^{ipm}-e^{-ipm}),
\end{equation}
\begin{equation}
\label{4.12a} C_{B'_{m}}=C'(e^{ipm}-z^{2}e^{-ipm}).
\end{equation}

Substituting Eqs.(\ref{4.11}-\ref{4.12a}) into
Eqs.(\ref{4.1}-\ref{4.4}) we obtain
\begin{equation}
\label{4.13}
(E-\Delta)\left(e^{ipm}-z^{2}e^{-ipm}\right)A+\left[\gamma_{0}\left(e^{ip(m-1)}-e^{-ip(m-1)}\right)+g_{k}\left(e^{ipm}-e^{-ipm}\right)\right]C+
\end{equation}
$$+\gamma_{1}\left(e^{ipm}-e^{-ipm}\right)C'=0,$$
\begin{equation}
\label{4.14}
\left[\gamma_{0}\left(e^{ip(m+1)}-z^{2}e^{-ip(m+1)}\right)+g_{k}\left(e^{ipm}-z^{2}e^{-ipm}\right)\right]A+(E+\Delta)\left(e^{ipm}-e^{-ipm}\right)C=0,
\end{equation}
\begin{equation}
\label{4.13a}
(E-\Delta)\left(e^{ipm}-e^{-ipm}\right)C'+\left[\gamma_{0}\left(e^{ip(m-1)}-z^{2}e^{-ip(m-1)}\right)+g_{k}\left(e^{ipm}-z^{2}e^{-ipm}\right)\right]A'+
\end{equation}
$$+\gamma_{1}\left(e^{ipm}-z^{2}e^{-ipm}\right)A=0,$$
\begin{equation}
\label{4.14a}
\left[\gamma_{0}\left(e^{ip(m+1)}-e^{-ip(m+1)}\right)+g_{k}\left(e^{ipm}-e^{-ipm}\right)\right]C'+(E+\Delta)\left(e^{ipm}-z^{2}e^{-ipm}\right)A'=0.
\end{equation}

This homogenous system of equations has a solution only if the
following conditions are fulfilled:
$$
\left[E^{2}-\Delta^{2}-\gamma_{1}(E+\Delta)-\left(\gamma_{0}e^{-ip}+g_{k}\right)\left(\gamma_{0}e^{ip}+g_{k}\right)\right]e^{2ipm}+
$$
$$z^{2}\left[E^{2}-\Delta^{2}-\gamma_{1}(E+\Delta)-\left(\gamma_{0}e^{-ip}+g_{k}\right)\left(\gamma_{0}e^{ip}+g_{k}\right)\right]e^{-2ipm}-$$
\begin{equation}
-(E^{2}-\Delta^{2}-\gamma_{1}(E+\Delta))(z^{2}+1)+\left(g_{n}+\gamma_{0}e^{ip}\right)^{2}+z^{2}\left(g_{n}+
\gamma_{0}e^{-ip}\right)^{2}=0,
\end{equation}
or
$$
\left[E^{2}-\Delta^{2}+\gamma_{1}(E+\Delta)-\left(\gamma_{0}e^{-ip}+g_{k}\right)\left(\gamma_{0}e^{ip}+g_{k}\right)\right]e^{2ipm}+
$$
$$
z^{2}\left[E^{2}-\Delta^{2}+\gamma_{1}(E+\Delta)-\left(\gamma_{0}e^{-ip}+g_{k}\right)\left(\gamma_{0}e^{ip}+g_{k}\right)\right]e^{-2ipm}-$$
\begin{equation}
-(E^{2}-\Delta^{2}+\gamma_{1}(E+\Delta))(z^{2}+1)+\left(g_{k}+\gamma_{0}e^{ip}\right)^{2}+z^{2}\left(g_{k}+\gamma_{0}e^{-ip}\right)^{2}=0.
\end{equation}

The coefficient of $e^{\pm 2pm}$ terms and the constant term have to
be equal to zero. Thus, we obtain the energy spectrum
\begin{equation}
\label{4.16} E_{1,2}=\frac{\gamma_{1}}{2}\pm
\sqrt{\gamma_{0}^{2}+2\gamma_{0}g_{k}\cos(p)+g_{k}^{2}+\left(\frac{\gamma_{1}}{2}+\Delta\right)^{2}},
\end{equation}
\begin{equation}
\label{4.16a} E_{3,4}=-\frac{\gamma_{1}}{2}\pm
\sqrt{\gamma_{0}^{2}+2\gamma_{0}g_{k}\cos(p)+g_{k}^{2}+\left(\frac{\gamma_{1}}{2}-\Delta\right)^{2}}.
\end{equation}
and the equation which gives the longitudinal wave number $p$ is
\begin{equation}
\label{4.18} \sin\left[p
M\right]+\frac{g_{k}}{\gamma_{0}}\sin\left[p(M+1)\right]=0.
\end{equation}
For $M\gg 1$ Eq.(\ref{4.18}) can be written as
\begin{equation}
\sin\left[pM\right]=0.
\end{equation}
The solution is given by
\begin{equation}
p=\frac{2\pi}{M}l.
\end{equation}
Substituting this solution into Eq.(\ref{4.16},\ref{4.16a}) we get
the energy spectrum  for the graphene bilayer with the periodic boundary
condition along the y-axis.

\section{Edge states of graphene bilayer}

Now we are interested in the edge state of the graphene bilayer. This
solution can be obtained in the form $p=\pi + i\eta$~\cite{Klein}.
We get the following equation for $\eta$:
\begin{equation}
\label{4.23} \sinh\left[\eta
M\right]-\frac{g_{k}}{\gamma_{0}}\sinh\left[\eta(M+1)\right]=0.
\end{equation}
The edge state can exist when the condition
\begin{equation}
\label{4.23a} |2\cos\left(k_{x}a/2\right)|<\frac{1}{1+1/M},
\end{equation}
is fulfilled. The energy spectrum of a state like that is given as

\begin{equation}
\label{4.24} E_{1,2}=\frac{\gamma_{1}}{2}\pm
\sqrt{\gamma_{0}^{2}-2\gamma_{0}g_{k}\cosh(\eta)+g_{k}^{2}+\left(\frac{\gamma_{1}}{2}+\Delta\right)^{2}},
\end{equation}
\begin{equation}
\label{4.24a} E_{3,4}=-\frac{\gamma_{1}}{2}\pm
\sqrt{\gamma_{0}^{2}-2\gamma_{0}g_{k}\cosh(\eta)+g_{k}^{2}+\left(\frac{\gamma_{1}}{2}-\Delta\right)^{2}}.
\end{equation}

For big enough $M$ the solution of Eq.(\ref{4.23}) can be expressed
in the form~\cite{Klein}
\begin{equation}
\label{4.26} \eta
=\ln\left[c_{k}+\frac{1-c_{k}^{2}}{c_{k}^{2M+1}}\right],
\end{equation}
where $1/c_{k}=|2\cos\left(\frac{k a}{2}\right)|$. We denote
$k_{x}=k$. From Eq.(\ref{4.26}) we have
\begin{equation}
\cosh\eta
\approx\frac{1+c_{k}^{2}}{2c_{k}}-\frac{\left(c_{k}^{2}-1\right)^{2}}{2c_{k}^{2M+3}},
\end{equation}
and so

\begin{equation}
\label{4.27} E_{1,2}=\frac{\gamma_{1}}{2}\pm
\sqrt{\gamma_{0}^{2}\frac{\left(c_{k}^{2}-1\right)^{2}}{c_{k}^{2M+4}}+\left(\frac{\gamma_{1}}{2}+\Delta\right)^{2}},
\end{equation}

\begin{equation}
\label{4.27a} E_{3,4}=-\frac{\gamma_{1}}{2}\pm
\sqrt{\gamma_{0}^{2}\frac{\left(c_{k}^{2}-1\right)^{2}}{c_{k}^{2M+4}}+\left(\frac{\gamma_{1}}{2}-\Delta\right)^{2}}.
\end{equation}

Now we assume, similarly as in~\cite{guinea}, that
$\gamma_{1}> 2\Delta$ and also it is assumed that the width of
the graphene bilayer is big enough and the following condition is
fulfilled:
\begin{equation}
\gamma_{1}\gg
\gamma_{0}^{2}\frac{\left(c_{k}^{2}-1\right)^{2}}{c_{k}^{2M+4}}.
\end{equation}
The bands are given by
\begin{equation}
\label{4.28}
E_{1}(k)=\gamma_{1}+\Delta+\frac{\gamma_{0}^{2}}{\gamma_{1}+2\Delta}\frac{\left(c_{k}^{2}-1\right)^{2}}{c_{k}^{2M+4}},
\end{equation}
\begin{equation}
\label{4.28a}
E_{2}(k)=-\Delta-\frac{\gamma_{0}^{2}}{\gamma_{1}+2\Delta}\frac{\left(c_{k}^{2}-1\right)^{2}}{c_{k}^{2M+4}},
\end{equation}
\begin{equation}
\label{4.29}
E_{3}(k)=-\Delta+\frac{\gamma_{0}^{2}}{\gamma_{1}-2\Delta}\frac{\left(c_{k}^{2}-1\right)^{2}}{c_{k}^{2M+4}},
\end{equation}
\begin{equation}
\label{4.28}
E_{4}(k)=-\gamma_{1}+\Delta-\frac{\gamma_{0}^{2}}{\gamma_{1}-2\Delta}\frac{\left(c_{k}^{2}-1\right)^{2}}{c_{k}^{2M+4}}.
\end{equation}
We are interested in the $E_{2}$($E_{3}$) band which is the
valence(conductance) band of the edge states. The minimum of the
$E_{2}$ band is
\begin{equation}
E_{2,min}=-\Delta -\frac{2}{e^{2}
M}\frac{\gamma_{0}^{2}}{\gamma_{1}+2\Delta},
\end{equation}
where it was used that $e^{x}=(1+x/M)^{M}$ for $M\rightarrow \infty
$ and also Eq.(\ref{4.23a}). We found $E_{2,max}=-\Delta$. Similarly
for the $E_{3}$ band

\begin{equation}
E_{3,max}=-\Delta +\frac{2}{e^{2}
M}\frac{\gamma_{0}^{2}}{\gamma_{1}-2\Delta},
\end{equation}
and $E_{3,min}=-\Delta$. We can see that the width of the band is
inversely proportional to the width of the bilayer.

The density of states can be expressed in the form
\begin{equation}
N(E)=\frac{L}{2\pi}\frac{1}{\frac{dE}{dk}},
\end{equation}
where $L$ is the length of the bilayer in the $x$ direction. We get
the density of the state in the region $E_{3,min}-E_{3,max}$ in the
vicinity of the energy $E=-\Delta$ in the form
\begin{equation}
N(E)=\frac{L}{2\pi a
M\left(E+\Delta\right)^{\frac{2M+1}{2M+2}}\left(\gamma_{0}^{2}/(\gamma_{1}-2\Delta)\right)^{1/2(M+1)}}.
\end{equation}
The density of the state in the region $E_{2,min}-E_{2,max}$ in the
vicinity of the energy $E=-\Delta$ has the form
\begin{equation}
N(E)=\frac{L}{2\pi a
M\left(E+\Delta\right)^{\frac{2M+1}{2M+2}}\left(\gamma_{0}^{2}/(\gamma_{1}+2\Delta)\right)^{1/2(M+1)}}.
\end{equation}
Both these densities of the states have a singularity at the energy
$E=-\Delta$. The strength of the singularity depends also on the
width of the bilayer. The width of the bilayer is characterized by
the parameter $M$. The density of the state of the
$E_{4}(k)(E_{1}(k))$ band is the same as the $E_{3}(k)(E_{2}(k))$
band.

\section{Conclusion}

In the presented paper the electronic spectra of the zig-zag bilayer
strip was studied analytically. We get for big enough $M$ that the
electronic spectra of the graphene bilayer strip are similar to the
spectra of the graphene bilayer with the periodic boundary condition.
Because of the boundary we also get edge states. It was shown
that the width of the edge state band is inversely proportional to
the width of the bilayer strip which is characterized by the
parameter $M$. So for big enough $M$ we get partly flat bands of the
edge states. The density of states at the Fermi level has a
singularity which also depends on the width of the bilayer strip.
There is asymmetry in the DOS at the Fermi energy, similarly to
the electron-hole asymmetry in the bilayer graphene~\cite{Falko}. This
asymmetry is caused by the parameter $\Delta$ which describes the
difference in the site energy of the atoms at the sites
$A_{i},B_{i}^{'}$ and the atoms at the sites $B_{i},A_{i}^{'}$.

\leftline{\bf Acknowledgment} \vspace{5mm}
The work was supported by the Slovak Academy of Sciences in the
framework of CEX NANOFLUID, and by the Science and Technology
Assistance Agency under Contract No. APVV 0509-07 , 0171 10,  VEGA
Grant No. 2/0069/10 and Ministry of Education Agency for Structural
Funds of EU in frame of project 26220120021.


\begin{thebibliography}{99}

\bibitem{Saito}Saito R,Dresselhaus G and Dresselhaus M S 1998 \textit{Physical Properties of Carbon Nanotubes}
(London:Imperial Colledge Press)
\bibitem{Wal}Wallace P R 1947 Phys.Rev. \textbf {71} 622
\bibitem{Carter} Carter J L and Krumhansl J A 1953 J.Chem.Phys. \textbf {21}
2238
\bibitem{Slon}Slonczewski J C and Weiss P R 1958 Phys.Rev. \textbf {109}
272
\bibitem{Novoselov} Novoselov K.S.,Geim A K,Morozov S V,Jiang D,Zhang Y,Dubonos S V,Grigoreva I V and Firsov A A
2004 Science \textbf {306} 666
\bibitem{pudlak} Pudlak M and Pincak R 2009 Phys.Rev. A \textbf {79}
033202
\bibitem{Fujita} Fujita M, Wakabayashi K, Nakada K and Kusakabe K 1996 J.Phys.Soc.Jpn. \textbf {65} 1920
\bibitem{Hod}Hod O and Scuseria G E 2008 ACS Nano \textbf {2} 2243
\bibitem{NOVA}Pincak R,Pudlak M and Smotlacha J in: 2012 \textit{Carbon Nanotubes:Synthesis,Properties and Applications}
(Nova Science Publisher,NY ) in press
\bibitem{Li} Li W and Tao R  2012 J.Phys.Soc.Jpn. \textbf {81} 024704
\bibitem{Wak}K.Wakabayashi,K.Sasaki,T.Nakanishi and T.Enoki 2010
Sci.Technol.Adv.Mater \textbf {11} 054504
\bibitem{Klein}D.J.Klein 1994 Chem.Phys.Lett.  \textbf{217} 261
\bibitem{guinea}F.Guinea,A.H.Castro Neto and N.M.R.Peres 2006 Phys.Rev. B  \textbf{73} 245426
\bibitem{Falko}Mucha-Kruczy\'{n}ski M,McCann and Fal\u{ }ko 2010 Semicond. Sci. Technol. \textbf{25}
033001


\end{thebibliography}
\end{document}